\documentclass[conference]{IEEEtran}
\IEEEoverridecommandlockouts
\usepackage{cite}
\usepackage{amsmath,amssymb,amsfonts}
\usepackage{algorithm}
\usepackage{algorithmic}
\usepackage{graphicx}
\usepackage{textcomp}
\usepackage{xcolor}
\usepackage{siunitx}
\def\BibTeX{{\rm B\kern-.05em{\sc i\kern-.025em b}\kern-.08em
    T\kern-.1667em\lower.7ex\hbox{E}\kern-.125emX}}
    
\def\R{\mathbb R}
\begin{document}

\title{
    Fine Grained Insider Risk Detection
}

\newcommand{\gauthor}[4]{
    \IEEEauthorblockN{
        #1
    }
    \IEEEauthorblockA{
       \small
        #4@google.com
    }
}

\author{
    \gauthor{Birkett Huber}{Brainauth}{Sunnyvale, USA}{bthuber}
    \and
    \gauthor{Casper Neo}{Brainauth}{New York, USA}{cneo}
    \and
    \gauthor{Keiran Sampson}{Insider Task Force (ITF)}{Sydney, Australia}{hpy}
    \and
    \gauthor{Alex Kantchelian}{Brainauth}{New York, USA}{akant}
    \and
    \gauthor{Brett Ksobiech}{SIRDE}{New York, USA}{bksobiech}
    \and
    \gauthor{Yanis Pavlidis}{Brainauth}{Sunnyvale, USA}{ypavlidis}

}

\maketitle

\begin{abstract}

We present a method to detect departures from business-justified workflows among support agents.
Our goal is to assist auditors in identifying agent actions that cannot be explained by the activity within their surrounding context,
where normal activity patterns are established from historical data.
We apply our method to help audit millions of actions of over three thousand support agents.

We collect logs from the tools used by support agents and construct a bipartite graph of \textit{Actions} and \textit{Entities}
representing all the actions of the agents, as well as background information about entities.
From this graph, we sample subgraphs rooted on security-significant actions taken by the agents.
Each subgraph captures the relevant context of the root action in terms of other actions, entities and their relationships.
We then prioritize the rooted-subgraphs for auditor review using feed-forward and graph neural networks,
as well as nearest neighbors techniques.
To alleviate the issue of scarce labeling data, we use contrastive learning and domain-specific data augmentations.

Expert auditors label the top ranked subgraphs as ``worth auditing" or ``not worth auditing" based on the company's business policies.
This system finds subgraphs that are worth auditing with high enough precision to be used in production.

\end{abstract}

\section{Introduction}

The U.S. Cybersecurity \& Infrastructure Security Agency (CISA) defines insider threat  \cite{cisa-it} as
“the threat that an insider will use their authorized access,  wittingly  or unwittingly, to do harm.”
Insider risk comes with potentially large financial impacts to companies,
costing on average \$16.2M per incident \cite{Ponemon}.
This system focuses on insider threats from support agents.

Support agents handle arbitrary support request tickets.
They must manage tickets appropriately in a ticket management system
and resolve the tickets, potentially by using their data access tools to read or modify sensitive data.
Tickets require human interpretation and may be routed between agents.
Their ability to route tickets and access sensitive data make them significant potential insider threats.

Previous work on insider risk detection,
including those on file access \cite{gates2014},
database access \cite{database_intrusion_detection_2010},
and general data access logs \cite{camlis23},
try to predict future (user, resource) access pairs from historical pairs.
Such systems are sub-optimal for detecting insider risk amongst support agents:
Which `resource' an agent should access depends on which tickets they
should be working on;
which, in turn, depends on variables extrinsic to the system;
and cannot be known ahead of time.
We call such systems `coarse grained' for neglecting within-workflow access justifications.

In contrast, we present a fine grained approach that
detects insider risk by finding departures from expected workflows. 
Whether an action is justified is determined by
other actions taken shortly before the access itself.
We use the term \textit{workflow} to encompass all activity related to the execution of a job function,
performed by a computer or human, particularly those that are logged by the tools used in the job function.

Specifying workflows in detail as hard-coded rules is difficult and toilsome
because a company's services and internal tooling change over time,
inevitably changing how support agents \textit{should} do their jobs,
as well as the logs describing how they \textit{did} do their jobs.
To avoid enumerating workflows, we use machine learning and learn them from the data.

\section{Methodology}
The logs produced by the support agents' tools are too low level to be individually audited for legitimacy.
In the course of their work, security analysts connect agents' activities across disparate systems by correlating low-level activity logs using timestamps or entity identifiers.

Inspired by their process, we contextualize logs by parsing them into a graph.
We extract subgraphs, rooted on potentially sensitive actions,
then prioritize them for review.
Security analysts then audit these subgraphs in order of priority. 

\subsection{Graph Construction}

We build a bipartite graph and name its partitions \textit{Actions} and \textit{Entities}.
Entities are identifiers that persist over time, including support agent usernames, account ids, and ticket ids.
Actions are temporally localized aggregations of logs with a categorical type.
A support agent commenting on a ticket is modeled as an action, with type `TicketManagement.Reply'.
That action is connected to the agent's entity and a ticket entity.
Edges connect actions and entities, and are annotated with a categorical \textit{relationship}.

We store our graph on disk as a large set of action records.
Each Action record has a unique \textit{id}, a string \textit{type}, \textit{start} time, \textit{end} time, and a set of \textit{references}.
A \textit{reference} has three string fields, \textit{entity type}, \textit{entity id}, and \textit{relationship}.
Note that entities are replicated everywhere they are referenced.

We select actions that have types indicating potentially sensitive activity (for example, `DataTool.Query')
and consider them single-node rooted subgraphs.
We expand those subgraphs using a breadth first traversal of the main graph.
For each of $T\in\mathbb N$ traversal steps,
for each entity that was not previously traversed,
and for each action type;
we add to the subgraph the $M\in\mathbb N$ actions 
who's start times are closest to the root action's start time,
who that reference that entity, are of that type, and were not previously added.

\begin{figure}[!htb]
    \centering
    \includegraphics[width=\linewidth]{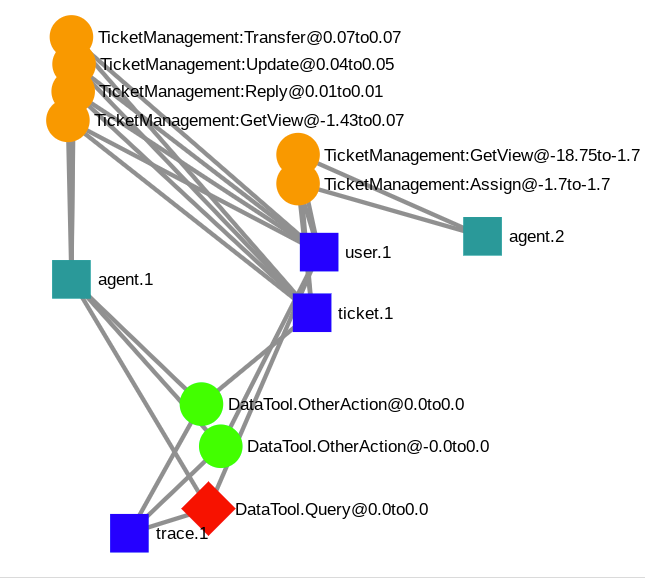}
    \caption{
        This subgraph is rooted on a \textbf{DataTool.Query} typed action (red diamond),
        which indicates \textbf{agent.1} queried \textbf{user.1}'s data.
        The \textbf{TicketManagement} tool recorded \textbf{ticket.1} pertains to \textbf{user.1},
        that \textbf{agent.1} viewed \textbf{ticket.1} 1.43 hours before taking the action, and
        that they transferred the ticket to themself 0.07 hours before making the query.
        We can also see that \textbf{ticket.1} was previously viewed by another agent, \textbf{agent.2},
        starting 18.75 hours before the root action, and assigned the ticket to themself 1.7 hours before the root action.
        Note that the relationships on each edge are omitted for clarity.
    }
    \label{fig:exampleSubgraph}
\end{figure}

$M$ and $T$ control the size of our subgraphs.
Action types vary in temporal density, and
selecting the $M$ closest actions by type stops the graph from being cluttered by noisy action types.
As we traverse further in graph distance, entities and actions become less relevant to the root action.
In this work, we take $T=2$ (no Entity type is allowed to be traversed more than two steps from the root) and $M=10$.
We also disable traversal through certain entity types after the first traversal.
See Figure \ref{fig:exampleSubgraph} for an example rooted-subgraph.

\subsection{Ranking Techniques}

Given sufficient labeled data, e.g. thousands of true positive examples, we would directly train a classifier to identify what's worth auditing.
However, due to high security analyst costs, gathering sufficient data is infeasible in our domain.
We present two ranking methods that require fewer labels.

\subsubsection{Nearest Neighbor (NN)}
Our first ranking technique is to rank subgraphs by their distance to a small set of interesting rooted-subgraphs, $I$.
In our evaluation, $|I|=2$.
Let $F$ be a function that maps rooted subgraphs, $N$, to the unit sphere in $\R^n$.
We define our distance function over graphs to be  $d_F(x, y) =\|F(x) - F(y)\|_2$.
Using this distance, we audit the $k$ closest subgraphs to $I$ from $N\setminus I$.

\subsubsection{Synthetic Mutation Rank (SMR)}
Our second ranking method exploits domain knowledge to find rooted subgraphs that should be interesting.
We create mutations that make a natural subgraph, $n\in N$, look more like un-expected (interesting) behavior.
We train a feed forward binary classifier to distinguish the embeddings of natural and mutated subgraphs.
Using this classifier, the top $k$ most seemingly-mutated natural subgraphs are selected for auditing.

\subsection{Generating Embeddings}
This section discusses our selection of $F$. We experiment with both hand-crafted features and learned ones.

\subsubsection{Handcrafted Embeddings}

As a simple starting point, we count the actions sharing either a user, an agent, or both with the central action, along with the earliest and latest start times for such actions, broken down by action type.
We rescale all these counts and relative time offsets with a signed-log, $x \mapsto sign(x)\log(1+|x|)$, and arrange them as a vector.

\subsubsection{Graph Neural Network (GNN) Embeddings}

Due to job specialization, support agents tend apply the same workflow repeatedly,
while different agents have different kinds of workflows.
We trained a self-supervised convolutional graph embedding network, $F_g$, to embed rooted subgraphs such that
the embeddings of a pairs of rooted-subgraphs are close if and only if their root actions were sampled from the same support agent on the same day.
Intuitively, this teaches the model to learn the aspects of subgraphs that differ between people and thereby different workflow types, while ignoring aspects that correspond to variations of same workflow.

Let $S_{F_g}((a,b))=\|F_g(a) - F_g(b)\|_2$ be the embedding distance of a pair of subgraphs.
Let $N$ be the distribution of rooted-subgraph pairs where the root actions correspond correspond to the same agent and the same day;
and $P$ be the distribution of rooted-subgraph pairs, where both items in the pair are independently sampled from the uniform distribution over all rooted-subgraphs.
Let $\phi\in(0, 1)$, $y\sim \text{Bernoulli}(\phi)$, and $z\sim D$ be a random pair that may be sampled from $P$ or $N$, depending on $y$.
Particularly, let $(z|y=0)\sim N$, and $(z|y=1)\sim P$.
Finally, let $L$ be the Huber loss, \cite{huber64}.
We train a model to optimize:
$$
\newcommand{\expected}[2]{\underset{#1}{\mathbb E}#2}
\min_{F_g}\expected{z\sim D}{L\Big((-1)^y S_{F_g}(z) \Big) }
$$

\section{Evaluation}
We evaluated one week of data, which contains over 95,000 rooted subgraphs describing actions of around 3100 active support agents,
and measure the system's precision from the perspective of an auditor.
To evaluate each method, two auditors rate each of the top $k$ rooted-subgraphs as “worth auditing” or “not worth auditing”
in that the support agent didn't follow expected guidelines or the tools didn't record things appropriately.
Both of these situations represent areas of improvement for the company. We present the number of subgraphs audited, $k$;
the number of those deemed worth auditing, $w$; and a Bayesian 90\% credible interval of precision.


\begin{center}
\newcommand{\hz}[0]{\hphantom{0}} 
\begin{tabular}{ |c||c|c|c| } 
 \hline
 Method         & $k$   & $w$   & Precision \% \\
 \hline \hline
 NN, Handcrafted    & 50    & 38    & 64.7  - 84.2  \\
 NN, GNN            & 50    & 38    & 64.7  - 84.2 \\
 SMR, Handcrafted   & 50    & 32    & 52.3  - 74.0  \\
 SMR, GNN           & 50    & 29    & 46.3  - 68.7 \\
 \hline
 Random             & 50    & \hz1  &\hz0.7 - \hz9.0 \\
 \hline
\end{tabular}
\end{center}
We find that all of our techniques significantly outperform the baseline of random audits.

Our NN ranking technique slightly outperforms SMR.
NN is easier to implement than SMR as the latter requires a domain expert to design mutations.
However, NN requires a set of interesting subgraphs 
and may limit future findings to be similar to those previously found.
While both handcrafted and GNN-based embeddings resulted in $w=38$, the particular subgraphs they found are different.

Handcrafted embeddings slightly outperform the GNN's with SMR ranking, but they are equivalent for NN ranking.
The GNN embeddings are more complex to create than handcrafted embeddings but the latter were constructed with domain
specific knowledge, and the GNN relies on the relatively generic hypothesis of workflow-agent affinity.

\section{Conclusion}

We presented a family of methods to model workflows and prioritize them for
security audits that do not require large amounts of labeled data and rely on
varying amounts of expert domain knowledge.
At time of publication, we are actively productionizing this system for continuous use.

In the future, we intend to generalize this work beyond support agent workflows.
To do this at scale, we need to automate the aspects of our method that require per-domain engineering.
Recent research suggests LLMs may be used to parse logs into Actions and Entities \cite{jiang2024lilac}.
Additionally, we want to automate selection of graph construction parameters.

\bibliographystyle{IEEEtran}
\bibliography{bibliography}

\begin{thebibliography}{10}
\providecommand{\url}[1]{#1}
\csname url@samestyle\endcsname
\providecommand{\newblock}{\relax}
\providecommand{\bibinfo}[2]{#2}
\providecommand{\BIBentrySTDinterwordspacing}{\spaceskip=0pt\relax}
\providecommand{\BIBentryALTinterwordstretchfactor}{4}
\providecommand{\BIBentryALTinterwordspacing}{\spaceskip=\fontdimen2\font plus
\BIBentryALTinterwordstretchfactor\fontdimen3\font minus
  \fontdimen4\font\relax}
\providecommand{\BIBforeignlanguage}[2]{{%
\expandafter\ifx\csname l@#1\endcsname\relax
\typeout{** WARNING: IEEEtran.bst: No hyphenation pattern has been}%
\typeout{** loaded for the language `#1'. Using the pattern for}%
\typeout{** the default language instead.}%
\else
\language=\csname l@#1\endcsname
\fi
#2}}
\providecommand{\BIBdecl}{\relax}
\BIBdecl

\bibitem{cisa-it}
``Defining insider threats,''
  \url{https://www.cisa.gov/topics/physical-security/insider-threat-mitigation/defining-insider-threats},
  accessed 2023-12-06.

\bibitem{Ponemon}
\BIBentryALTinterwordspacing
``2023 cost of insider threat global report. ponemon institute.'' [Online].
  Available:
  \url{https://www.dtexsystems.com/resource-ponemon-insider-risks-global-report/}
\BIBentrySTDinterwordspacing

\bibitem{gates2014}
C.~Gates, N.~Li, Z.~Xu, S.~N. Chari, I.~Molloy, and Y.~Park, ``Detecting
  insider information theft using features from file access logs,'' in
  \emph{19th European Symposium on Research in Computer Security
  ({ESORICS}}.\hskip 1em plus 0.5em minus 0.4em\relax Springer, 2014, pp.
  383--400.

\bibitem{database_intrusion_detection_2010}
S.~Mathew, M.~Petropoulos, H.~Q. Ngo, and S.~Upadhyaya, ``A data-centric
  approach to insider attack detection in database systems,'' in \emph{Recent
  Advances in Intrusion Detection}, S.~Jha, R.~Sommer, and C.~Kreibich,
  Eds.\hskip 1em plus 0.5em minus 0.4em\relax Berlin, Heidelberg: Springer
  Berlin Heidelberg, 2010, pp. 382--401.

\bibitem{camlis23}
\BIBentryALTinterwordspacing
G.~Gelven and S.~Strum, ``Graph-based user-entity behavior analytics for
  enterprise insider threat detection.'' [Online]. Available:
  \url{https://www.camlis.org/grant-gelven-2023}
\BIBentrySTDinterwordspacing

\bibitem{huber64}
\BIBentryALTinterwordspacing
P.~J. Huber, ``{Robust Estimation of a Location Parameter},'' \emph{The Annals
  of Mathematical Statistics}, vol.~35, no.~1, pp. 73 -- 101, 1964. [Online].
  Available: \url{https://doi.org/10.1214/aoms/1177703732}
\BIBentrySTDinterwordspacing

\bibitem{jiang2024lilac}
Z.~Jiang, J.~Liu, Z.~Chen, Y.~Li, J.~Huang, Y.~Huo, P.~He, J.~Gu, and M.~R.
  Lyu, ``Lilac: Log parsing using llms with adaptive parsing cache,'' 2024.

\bibitem{tfgnn}
\BIBentryALTinterwordspacing
O.~Ferludin, A.~Eigenwillig, M.~Blais, D.~Zelle, J.~Pfeifer,
  A.~Sanchez{-}Gonzalez, W.~L.~S. Li, S.~Abu{-}El{-}Haija, P.~Battaglia,
  N.~Bulut, J.~Halcrow, F.~M.~G. de~Almeida, P.~Gonnet, L.~Jiang, P.~Kothari,
  S.~Lattanzi, A.~Linhares, B.~Mayer, V.~Mirrokni, J.~Palowitch, M.~Paradkar,
  J.~She, A.~Tsitsulin, K.~Villela, L.~Wang, D.~Wong, and B.~Perozzi,
  ``{TF-GNN:} graph neural networks in tensorflow,'' \emph{CoRR}, vol.
  abs/2207.03522, 2023. [Online]. Available:
  \url{http://arxiv.org/abs/2207.03522}
\BIBentrySTDinterwordspacing

\bibitem{gatv2}
S.~Brody, U.~Alon, and E.~Yahav, ``How attentive are graph attention
  networks?'' 2022.

\bibitem{vizier}
D.~Golovin, B.~Solnik, S.~Moitra, G.~Kochanski, J.~Karro, and D.~Sculley,
  ``Google vizier: A service for black-box optimization,'' in \emph{Proceedings
  of the 23rd {ACM} {SIGKDD} International Conference on Knowledge Discovery
  and Data Mining}.\hskip 1em plus 0.5em minus 0.4em\relax {ACM}, 2017, pp.
  1487--1495.

\end{thebibliography}

\section{Appendix}

\subsection{GNN Training}
Our model is trained using Tensorflow GNN\cite{tfgnn}.
The graph is convolved with a parameterized number of GATv2 attention \cite{gatv2} layers.
In each convolution, entities attend to the linked actions then actions attend to the linked entities.
Then, we initialize a global node as a function of the root action.
The global node attends to all actions and entities for a parameterized number of rounds.
Finally, the global node's embedding is L2 normalized and output.

We use Google Vizier \cite{vizier}, a black box optimization service, to select
hyperparameters. Let $F_{gh}$ be the GNN model trained with hyperparameters, $h$, chosen by Vizier.
Since the training objective is affected by hyperparameters, such as $\phi$ and the Huber loss parameters,
Vizier should not be tasked not tasked with minimizing validaiton loss.
Instead, we provide the objective of maximizing the cross entropy of $N$ to $P$ when pushed forward through $S_{F_{gh}}$,
$$
    \max_{h}\underset{S_{F_{gh}}(P)}{\mathbb E} -\log S_{F_{gh}}(N)
$$
We use the best model (according to this objective) found by Vizier in our evaluation.


\end{document}